# Expanding the Human Ancestry Ontology to include under-represented populations and ethnicities for broader utility in annotations

Anita Caron[1], Melek Chaouch[2], Zoë May Pendlington[1], Helen Parkinson[1], and Danielle Welter[3,*]

1. European Bioinformatics Institute (EMBL-EBI), Wellcome Genome Campus, Hinxton, UK.
2. Laboratory of BioInformatics, bioMathematics and bioStatistics (BIMS), LR16IPT09, Institut Pasteur de Tunis, University of Tunis El Manar, Tunis, Tunisia.
3. Luxembourg National Data Service, PNED GIE, 4362 Esch-sur-Alzette, Luxembourg.

*Correspondence: danielle.welter@lnds.lu

## Summary

Successful discovery, integration and reuse of data relies on the availability of rich, well-structured and machine-readable metadata to describe every aspect of the data, from sample sources to collection processes to experimental protocols. The use of standardised terminologies to express concepts in a harmonised fashion lies at the core of high-quality data annotation, increasing the FAIRness of the data, facilitating data integration and promoting reproducibility. Here, we describe the Human Ancestry Ontology (HANCESTRO), originally developed to improve standardised reporting of genetic ancestry genomic resources such as the NHGRI-EBI GWAS Catalog and the Human Cell Atlas through high-level population descriptors, and more recently expanded to include diverse and previously under-represented populations in genomics and genetics research. HANCESTRO provides a framework for population descriptors that includes both ancestry based on the analysis of genetic information and self-reported ethnicity, which is based on social and cultural factors that don't necessarily align with genetic populations. By enabling the accurate and interoperable representation of population-related data, it promotes inclusive, representative and reproducible science.

## Introduction

In the era of big data, high-quality, structured and interoperable data annotations are crucial for effective data findability and reuse. For biomedical research data, this includes the recording of experimental metadata from sample descriptions such as species, sex, tissue and sample collection and storage methods, to machine-readable experimental protocols, to data post-processing and storage information, all the way to the information reported in publications, both as free-text and as machine-readable formats. Ontologies play a key role in this challenge by providing a structured representation of domain-relevant concepts, including unique identifiers, standard definitions, synonyms, and relationships between concepts. Critically, the formal, logic-based relationships that distinguish an ontology from a flat controlled vocabulary or code list, such as hierarchical subsumption and defined properties between terms, enable computational reasoning over annotations, allowing queries and analyses to generalize or specialize across related concepts rather than being limited to exact string matches. The use of ontologies in data annotation not only reduces the variability inherent in unstructured text but also supports advanced data retrieval and data linkage, through search query expansions to broader and narrower concepts, and integration of data from different sources into knowledge graphs for extended querying and knowledge discovery. Here, we describe the latest developments of the Human Ancestry Ontology (HANCESTRO), including its restructured population-description framework, the introduction of reference populations for genetic similarity approaches, and its adoption across major biomedical resources.

In recent years, biomedical research, particularly genomics, has engaged in extensive collective soul-searching to identify and address pervasive bias and inequality in the representation of human study populations in research. Concerns have been raised about the misuse of genetic data to perpetuate outdated notions around race and ancestry[1–3]. A 2023 report[4] by the National Academies of Sciences, Engineering, and Medicine (NASEM) established a range of recommendations including avoiding typological thinking, improving transparency around the use of population descriptors and the reasoning behind it, and focusing on genetic similarity in preference to inherently simplistic group labels that fail to adequately reflect the continuum of genetic diversity.

With this wider context in mind, the ongoing development of an ontology for human ancestry may seem discordant with these recommendations. We acknowledge the importance of the NASEM report recommendations and have sought to incorporate them wherever possible. Critically, HANCESTRO is designed to support annotation rather than classification: its terms provide a standardised way to record population descriptors as they are reported in source data - self-identified ethnicity, cohort recruitment criteria, or genetically-defined reference panels - without asserting that any such grouping is biologically discrete, hierarchical, or temporally stable. The ontology's structure reflects this: ancestry, ethnicity, and geography-based terms are represented as parallel description frameworks rather than a single nested taxonomy, while the recent addition of reference population terms specifically operationalizes NASEM's recommendation to favour genetic similarity over ancestry-group labeling, giving researchers a standardized, machine-readable way to cite *which* reference panel was used, without HANCESTRO itself asserting that the resulting similarity groupings are natural or discrete population categories. HANCESTRO does not adjudicate which descriptor is appropriate for a given study; that determination remains the responsibility of researchers, guided by frameworks such as the NASEM report. The challenges of accurate, machine-readable data annotation nonetheless necessitate pragmatic, if imperfect, tools, and HANCESTRO aims to be one such tool, evolving as annotation needs and scientific understanding evolve. HANCESTRO does not compete with existing population-labelling systems but rather formally represents them, such as, for example, incorporating the 1000 Genomes Project[5] superpopulations as reference population terms, so that resources using different descriptor systems can be annotated within a single, interoperable framework.

HANCESTRO was developed in three major phases. It was first released in 2015 under the name ANCESTRO and the namespace "http://www.ebi.ac.uk/ancestro/", as a supporting component of the proposed standardized framework for representing ancestry data in genomics studies described in Morales et al[6]. The driver behind the initial development was the lack of coverage of relevant concepts in other ontologies and the problematic representation of available concepts that conflated race, ethnicity, and ancestry. During the second development phase, the ontology was renamed to HANCESTRO to emphasise its human-specific focus and migrated to the namespace "http://purl.obolibrary.org/obo/hancestro", while interoperability improvements were implemented by reusing a common upper-level structure.

The third development phase not only included the setup of a reusable development and release infrastructure but also extended the scope of the ontology beyond its original remit. While the initial version of the ontology focused exclusively on the concept of "ancestry category", with the supporting concept "ancestry status", and geography-based concepts for country, region and continent, new use cases, including the representation challenges discussed above, drove a major restructure in the following areas:

1. The geography-related part of the hierarchy (including countries, regions and continents) was migrated to a different philosophical parent in line with current best practices understanding in the biomedical ontologies community.
2. New frameworks for describing populations in terms of ethnicity and geographic location were put in place to address the evolving needs of user communities.
3. Reference populations were added to facilitate the implementation of NASEM report recommendations regarding the use of genetic similarity.

Since its initial release, HANCESTRO has been adopted by a number of projects and resources, with HANCESTRO terms now used to annotate ancestry and ethnicity in resources spanning data discovery and analysis platforms, single-cell atlases, and biosample archives, as well as being imported by several other biomedical ontologies.

In this paper, we focus on primarily the third development phase of HANCESTRO. Earlier development work is mentioned in the Methods section for context but not discussed extensively elsewhere.

| ***Info box 1: definitions for common concepts used in this paper*** | |
|---|---|
| Ancestry | An individual's background, origin or heritage. |
| Genetic ancestry | An individual's genetic heritage, inherited from their biological ancestors. Genetic ancestry can be inferred in terms of alignment or similarity with a defined reference population. Care should be taken not to rely exclusively on distinct labels, as these inherently hide the complexity of human ancestry. |
| Ethnicity | An individual's self-identified background based on cultural, linguistic, religious, tribal or other social determinants. Although ethnicity has no scientific basis, it may be a helpful supporting variable in some contexts. |
| Race | A sociopolitical classification based on outdated and subjective notions |
| | of innate biological similarity. This term should be avoided in any context, in particular in scientific research. |
| Population descriptor | A label for grouping individuals together according to a shared characteristic such as geography, ethnicity or genetic similarity. |

# Methods

## Ontology development

HANCESTRO is implemented in the Web Ontology Language (OWL)[7], a World Wide Web Consortium standard. To facilitate integration and interoperability with other ontologies, HANCESTRO is developed in line with the Open Biomedical Ontologies (OBO) development principles and best practices[8]. Through its namespace "http://purl.obolibrary.org/obo/hancestro", the ontology leverages the OBO Foundry Permanent URLs (PURLs) service to provide resolvable and persistent identifiers for all its terms.

OBO best practice also influenced building the ontology under the Basic Formal Ontology[9] (BFO) upper-level ontology classes to structure its hierarchy. While the use of a common upper structure, such as BFO, is essential for interoperability between ontologies, its philosophy-based terminology and highly technical definitions can be confusing for non-expert users. For this reason, HANCESTRO uses the ontology-level annotation property "has ontology root term" (IAO:0000700) to flag its conceptually most important root terms. This approach allows services such as the Ontology Lookup Service[10] (OLS) to display a simplified hierarchy to users.

In line with best practice, HANCESTRO reuses concepts from a number of ontologies, where these concepts are either outside the defined scope of HANCESTRO or where the source ontology is considered a greater authority on the definition of the concept. This includes all geographical concepts such as countries and continents, since HANCSTRO is not a domain authority on geography and these concepts are therefore not in the scope of HANESTRO. A full list of relevant ontologies from which HANCESTRO imports terms is provided in Table 1.

HANCESTRO defines 5 ontology root terms: (1) "ancestry category" (HANCESTRO:0004), (2) "ethnicity category" (HANCESTRO:0601), (3) "geography-based population category" (HANCESTRO:0602), (4) "reference population" (HANCESTRO:0632) and (5) "ancestry status" (HANCESTRO:0304). The first four of these are subclasses of "population" (OBI:0000181), which, in turn, is defined as a subclass of "material entity" (BFO:0000040) in its source ontology[11]. "Ancestry status" is defined as a subclass of "quality" (BFO:0000019). Although geographical concepts are essential to several of HANCESTRO's ontology design patterns, they are not part of the ontology's core scope and are therefore not flagged as root terms.

Finally, HANCESTRO also defines a range of design patterns[12], which represent standard relations between types of concepts in an ontology and which facilitate reasoning over and querying of ontology concepts.

| ***Table 1: A list of all ontologies from which HANCESTRO imports terms. It should be noted that this does not include transitive or incidental imports of terms that occur from these upstream ontologies via the import mechanism*** | | | |
|---|---|---|---|
| **Ontology** | **Number of imported terms** | **Reason for inclusion** | **Import mechanism** |
| Basic Formal Ontology[9] (BFO) ifomis.org/bfo/ | 8 | Interoperability with other ontologies | ODK default with SLME-BOT (Syntactic Locality Module Extractor for bottom module) |
| OBO Relations Ontology[13] (RO) https://oborel.github.io/ | 12 object properties | Reuse of object properties in line with best practice | ODK default |
| Ontology of Biomedical Investigations[11] (OBI) https://obi-ontology.org/ | 2 | Primary source of the terms "population" (OBI:0000181), which links the HANCESTRO scope to BFO, and "organization" (OBI:0000245) (no longer in use) in the OBO Foundry. | Custom import (ROBOT template - MIREOT) to minimise inclusion of unnecessary terms |
| African Population Ontology[14] (AfPO) https://github.com/h3abionet/afpo | 391 | Primary authoritative source for all populations for the continent of Africa. | Custom import (ROBOT template) to allow removal of overlapping terms imported into AfPO from HANCESTRO |
| Gazetteer[15] (GAZ) https://environmentontology.github.io/gaz/ | 2 terms, 259 database cross references | Primary source of terms "geographic location" (GAZ:00000448) and "continent" (GAZ:00000013) in the OBO Foundry; cross-reference of GAZ individual. | ODK component using ROBOT template for cross-reference. Stand-alone ROBOT template for hierarchy. |
| National Cancer Institute Thesaurus[16] OBO Edition (NCIT) https://ncithesaurus.nci.nih.gov/ncitbrowser/ | 1 | Primary source of the term "Country" (NCIT:C25464) in the OBO Foundry. | N/A, stand-alone ROBOT template for hierarchy. |
| DBpedia[17] https://www.dbpedia.org/ | 267 | Selected as source for geographical concepts due to extent and quality of available metadata and format interoperability with OWL during the earliest HANCESTRO development phase. | ODK component using ROBOT template |

## Geographical concepts in HANCESTRO

HANCESTRO's use of geographical concepts underwent several major evolutions. Early versions of the ontology classified most geographical concepts under the parent "material entity" (BFO:0000040), with the exception of countries, which were classified under "organization" (OBI:0000245), following the precedent set by other ontologies at the time. The concepts of "region" and "country", as well as the subclasses for region and continent were defined natively in HANCESTRO due to a lack of suitable concepts available for reuse from other ontologies, although the intention was always to eventually replace them, if possible. Countries were imported from DBPEDIA[17], as described in Table 1, due to the breadth and quality of the metadata available for these concepts, and the resource's general interoperability with OWL ontologies.

While Gazetteer[15] (GAZ) already existed prior to HANCESTRO's earliest development, it could not be reused directly for complexity reasons. During the v3 development phase, cross-references between geographical terms in HANCESTRO and GAZ terms were introduced for greater interoperability with resources that do use GAZ. At the same time, all geographical concepts were reclassified as subclasses of "immaterial entity" (BFO:0000141), in line with the now more common classification of geographic locations as immaterial entities, as this representation is most consistent with the definition of immaterial compared to material entities. The reclassification of country, region and continent concepts underwent multiple iterations, in part due to changes in upstream ontologies. The current version of HANCESTRO reuses the concepts of "continent" (GAZ:00000013) and Country (NCIT:C25464), both subsumed by "geographic location" (GAZ:00000448). Only "region" (HANCESTRO:0002) was maintained as a HANCESTRO-specific term as no suitable equivalent could be identified elsewhere. All continents and regions were migrated from HANCESTRO-specific concepts to their equivalent DBPEDIA terms, except for the region "Australia/New Zealand" (HANCESTRO:0051), for which no exact equivalent exists.

### Geographic design patterns

HANCESTRO defines the following design patterns for its geographic terms:

- Region part_of only continent
- Country located_in only region

These patterns are intended to facilitate geography-based querying within the ontology, e.g. in order to identify concepts linked, via another pattern, to countries located within a specific region of interest.

## Population descriptor frameworks

HANCESTRO's original scope covered only ancestry as defined in Morales et al., accordingly defining 16 equivalent ancestry categories and sub-categories. Specific population descriptors, usually linked to geographic locations, were classified under one of the main categories, depending on available evidence, again in line with the work described in Morales et al, e.g. "Indian" (HANCESTRO:0487), demonym of country India and classified as "South Asian" (HANCESTRO:0006). The category "undefined ancestry population" (HANCESTRO:0566) was added for population descriptors not covered in previous work and terms such as Cape Verdean (HANCESTRO:0504) that were identified in Morales et al but not allocated to a wider ancestry group. HANCESTRO also defined the quality term “ancestry status” (HANCESTRO:0304) with the child terms “admixed ancestry” (HANCESTRO:0306) and “genetically isolated ancestry” (HANCESTRO:0305), to allow for easy querying of descriptors matching these categories.
Emerging use cases to capture not only genetic ancestry but also self-reported ethnicity, as well as the recommendations of the NASEM report[4], drove the expansion of the ontology’s coverage to other types of population descriptors. including categories for ethnicity descriptors (“ethnicity category” (HANCESTRO:0601)), geography-based descriptors (“geography-based population category” (HANCESTRO:0602)) and reference populations (“reference population” (HANCESTRO:0632)). Alongside these new categories, two new quality terms were defined, “ethnicity descriptor” (HANCESTRO:0599) and “geographic descriptor” (HANCESTRO:0600).

## Reference populations

Reference populations consist of groups of individuals from a common geographic area and tribal or ethnolinguistic group whose DNA was sampled and sequenced as part of a large-scale genomic diversity project such as the 1000 Genomes Project[5], the Human Genome Diversity Project[18] (HGDP) or Simons Genome Diversity Project[19] (SGDP). Reference populations play a crucial role in human genomic data analysis, so to meet evolving user needs, HANCESTRO now includes the full list of reference populations defined in the aforementioned projects and listed in the International Genome Sample Resource[20] (IGSR, https://www.internationalgenome.org/data-portal/sample).
For each reference population in IGSR, a definition was created from the information provided, including the number of individuals, the recruitment location, the specific project under which the population was defined and the superpopulation to which the collecting project allocated the population. We also mapped each reference population to an ancestry category in HANCESTRO, based on the superpopulation defined by the source project and we defined “partial overlap” relationships between reference population terms and other HANCESTRO population descriptors, where these were available, in order to again indicate that an overlap may exist between individuals with a high degree of genetic similarity to a reference population and individuals self-identifying with or being referred to using that population descriptor.

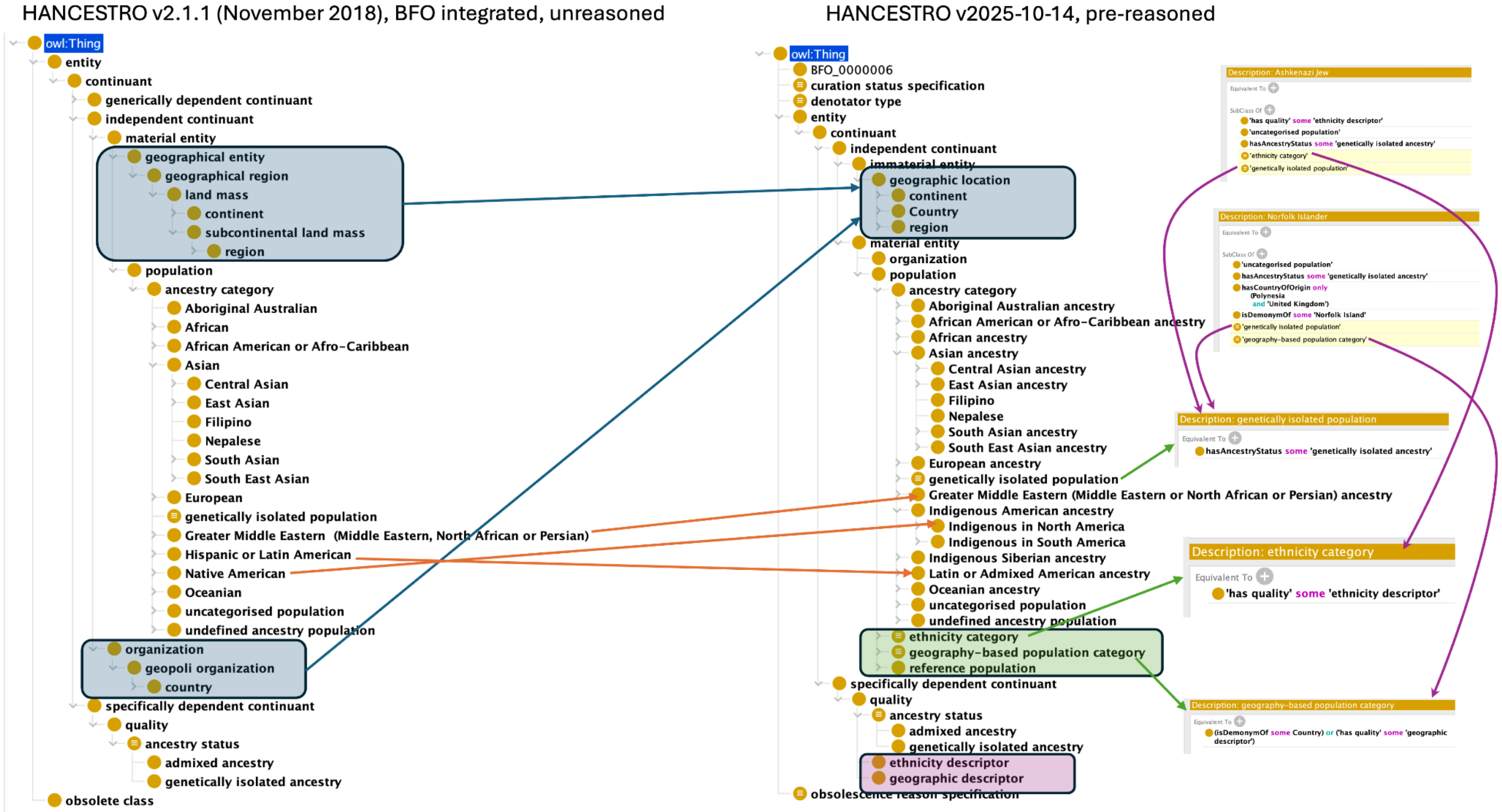

*Figure 1: HANCESTRO changes over time. The blue boxes and arrows highlight the evolution of the geographic concept classification. Orange arrows indicate ancestry groups that were renamed and/or moved since the initial release. Green and purple boxes highlight the new descriptors and descriptor qualifiers, respectively. The boxes on the right side illustrate the design patterns for the three equivalent classes as well as two examples, "Norfolk Islander" and "Ashkenazi Jew", to illustrate the impact of reasoning-based inference.*

## Population design patterns

HANCESTRO defines the following design patterns for its population descriptor terms:

- Population descriptors:
    - <descriptor> isDemonymOf some country
    - <descriptor> hasAncestryStatus some admixed ancestry
    - <descriptor> hasAncestryStatus some genetically isolated ancestry
    - <descriptor> has quality some geographic descriptor
    - <descriptor> has quality some ethnicity descriptor
- Equivalent classes:
    - Genetically isolated population equivalent to (hasAncestryStatus some genetically isolated ancestry)
    - Geography-based population category equivalent to (isDemonymOf some country) or (has quality some geographic descriptor)
    - Ethnicity category equivalent to (has quality some ethnicity descriptor)
- Reference populations:
    - <reference population> hasCountryOfOrigin some <country>
    - <reference population> partially overlaps some <descriptor>

The driving use case for these patterns is either to facilitate querying or to support automated grouping of terms without the need to directly assert multiple parent terms. As an example, early users of HANCESTRO expressed the need to easily identify genetically isolated populations. Rather than assert affected terms as child term of both the ancestry label to which they were initially attributed and the separate "genetically isolated population" (HANCESTRO:0290) group, the terms were instead annotated with the "<descriptor> hasAncestryStatus some genetically isolated ancestry" pattern to allow automated classification under HANCESTRO:0290 using an ontology reasoner.

## Infrastructure

HANCESTRO put in place a robust and sustainable development and release infrastructure used to build, validate, release and document the ontology. The implementation uses the Ontology Development Kit[21] (ODK), a versatile toolkit integrating several common ontology development libraries to facilitate the building, maintenance and standardisation of biomedical ontologies. A YAML-based config file defines every aspect of the ontology, from formats and namespace to import modules from external sources and components. Build and release processes are governed by a standard Make file, although custom extensions were necessary in a couple of scenarios, as described in Table 1.

The ODK also includes an out-of-the-box framework for setting up documentation sites, leveraging the GitHub Pages functionality. An ontology documentation wiki for HANCESTRO is available at https://ebispot.github.io/hancestro/.

During the third development phase, specifically during the integration with AfPO[14], HANCESTRO moved from HERMIT[22] to ELK[23] as its ontology reasoner, as the increased ontology size made HERMIT-based reasoning unscalable. Unlike HERMIT, ELK only supports a subset of OWL axioms (OWL2-EL[24] profile), which meant that some of the design patterns had to be adjusted to ensure reasoning remained consistent. For example, "hasAncestryStatus *only* genetically isolated ancestry" had to be changed to "hasAncestryStatus *some* genetically isolated ancestry". The ontology still contains a few axioms not supported by ELK but these are limited to the geographic location hierarchy and have minimal impact on reasoning, so the decision was made to keep these unchanged.

# Results

## A sustainable technical infrastructure

The migration from Protege ontology editing to the ODK framework enabled the introduction of standardised, well-documented editing, quality control and release processes. Ontology updates are now performed almost exclusively via template files rather than direct editing of the ontology source files, greatly improving transparency and reproducibility while minimising the risk of accidental changes in the source files. The ODK setup also allows the automatic creation of full release diffs (https://github.com/EBISPOT/hancestro/blob/main/src/ontology/reports/release-diff.md),

ensuring that each release is transparently documented without relying on editors to remember all the changes included in the release and manually document them.
The ODK framework documentation component allowed us to easily set up and maintain a HANCESTRO documentation site based on GitHub Pages. This wiki-style site now contains up-to-date documentation on HANCESTRO's background, its content, usage and editorial processes.

## Ontology metrics

Release v2025-10-14 of HANCESTRO contains a total of 1310 classes, of which 330 are defined directly in HANCESTRO.

***Table 2: Class count for HANCESTRO release v2025-10-14. *Note that some terms appear under multiple parents, so the cumulative class count is greater than the number of unique classes.***

| Concept | Class count* |
|---|---|
| Ancestry category | 952 |
| Ethnicity category | 464 |
| Geography-based population category | 235 |
| Reference population | 231 |
| Continent | 6 |
| Region | 22 |
| Country | 240 |

## Ancestry, geographic and ethnicity descriptors

HANCESTRO's initial scope focused on defining broad ancestry categories, largely aligned with major continental groups. More detailed terms mapped to these broad categories were derived from the detailed sample descriptions provided in genome-wide association studies (GWAS) publications curated by the GWAS Catalog, as described in Morales et al.[6]. Over time, additional descriptors were requested by other projects also using HANCESTRO in their data curation and annotations. Some of these terms did not fit into any of the defined ancestry categories, leading to a *de facto* allocation to the "undefined ancestry population" (HANCESTRO:0566) group. Even this term, however, falls under the root term of "ancestry category" (HANCESTRO:0004), which is defined as "Population category defined using ancestry informative markers (AIMs) based on genetic/genomic data". In order to accommodate population descriptors that are purely based on cultural, religious, tribal or other social affiliation, a separate root concept, "ethnicity category" (HANCESTRO:0601), was defined.

In line with the recommendations of the NASEM report[4], which encourage the use of a single descriptor type in a given context (Recommendation 7), we also created a similar grouping category for geographic descriptors ("geography-based population category" (HANCESTRO:0602)). Although we decided not to make ethnicity and geographic descriptors mutually exclusive on a technical level by making the two classes disjoint, every effort was made to ensure that the two groups are non-overlapping. Due to the hierarchical constraints of OWL ontologies, which dictate that all properties and characteristics of a class also apply to any of its descendants, it was in some cases necessary not to classify a term as a geography-based descriptor if it has child terms that are ethnicity descriptors. One such example is "Andalusian" (AfPO:0000355), a term imported from the African Population Ontology (AfPO), which subsumes "Tunisian Andalusian" (AfPO:0000363): the former is the demonym for the region of Andalusia in Spain, making it technically a geographic descriptor but the latter is a cultural descriptor referring to the descendants of Muslims expelled from Spain in the seventeenth century, who sought refuge in Tunisia[25]. As it is impossible to classify "Andalusian" as a geographic descriptor without applying the same "Tunisian Andalusian", therefore both terms are tagged as ethnicity descriptors in this case.

In a similar vein, we appended the suffix "ancestry" to the original ancestry categories from the GWAS ancestry framework in order to be able to create overlapping but distinct geography-based descriptors for some of the larger sub-continental regional groups, e.g. "South East Asian ancestry" (HANCESTRO:0007) vs "Southeast Asian" (HANCESTRO:0850). Although the suffix was applied to all the original ancestry categories, separate geography-based descriptors were created only for those terms commonly used in the US, UK and EU as super-national geographic descriptors in non-genetic contexts. These enable researchers to accurately annotate data, prevent loss of data, and comply with governmental and funding standards around data collection.

In its current implementation, "geography-based population category" is a completely flat list while "ethnicity category" inherits some localised hierarchies from the main ancestry groups. The majority of these originate in AfPO, which has a more nested hierarchy than most of HANCESTRO.

## African Population Ontology import

As previously discussed, HANCESTRO reuses classes from other ontologies, in line with ontology development best practice - most notably from the African Population Ontology (AfPO) (https://github.com/h3abionet/afpo). AfPO was developed as part of the H3ABioNet (the Pan-African Bioinformatics Network for H3Africa) and the AfriGen-D (African Genomic Data Hub) projects, making it the authoritative source on African populations in the biomedical domain. Unlike other ontologies from which HANCESTRO imports terms, AfPO covers a similar scope, although the two ontology projects differ in their initial approach to identifying and defining populations. AfPO started from linguistics resources to integrate population labels with anthropological information such as spoken languages, demographic data including geographic origin, and countries of residence, and finally genomic data, aligning multiple layers of evidence for grouping or distinguishing populations. While HANCESTRO has focused primarily on the genomic component and population labels reported in GWAS publications, there is sufficient alignment in the respective scopes of the two ontologies for HANCESTRO to defer to AfPO for all its African populations.
The AfPO developers in turn worked closely with the HANCESTRO team to harmonise the upper structure of the ontologies, implementing HANCESTRO's approach to geographic locations as well as HANCESTRO's "country of origin" pattern. All population terms created by AfPO are included in the HANCESTRO hierarchy underneath the continental ancestry term African (HANCESTRO:0010), with Central, Eastern, Southern and Western African categories nested under "Sub-Saharan African" (HANCESTRO:0011), a concept not recognised by AfPO but maintained in HANCESTRO for backwards compatibility with the original framework. Existing HANCESTRO terms such as Masaai (HANCESTRO:0593) or Somali (HANCESTRO:0518) were deprecated in favour of their AfPO equivalents.

## Reference populations

Reference populations from large-scale genomic diversity projects, including the 1000 Genomes Project, the Human Genome Diversity Project (HGDP) and Simons Genome Diversity Project (SGDP) were recently added to HANCESTRO to address new use cases related to human genomic data analysis. In particular, the latest guidance around ethical use of population descriptors[4] recommends defining study samples through the percentage of genetic alignment with a named reference population rather than the use of potentially imprecise or outdated descriptors based on cultural or geographic background.
By mapping each reference population to an ancestry category in HANCESTRO, we aimed to provide an interoperability layer between reference populations and population descriptors likely used in similar contexts. It is worth noting that these mappings do not imply that a reference population constitutes an exact synonym for the ancestry category it is mapped to, or that ancestry labels represent shorthand designations for reference panels that can be used interchangeably. The mappings are only intended to indicate that a degree of overlap may exist between individuals that have a high degree of genetic alignment with a reference population and individuals that self-identify with the related ancestry label.
For the majority of populations, the superpopulation allocated by the source project made it relatively straightforward to identify a suitable HANCESTRO category to express these mappings. The most challenging superpopulations to map were the "West Eurasia" and "Central Asia and Siberia" superpopulations defined by the SGDP as these did not map directly to any existing HANCESTRO concepts. For "West Eurasia", we used Extended Data Figure 3 (K=9 to K=12) in Mallick et al.[19] combined with existing ancestry group mappings in HANCESTRO to determine the most appropriate ancestry group for each individual reference population. For "Central Asia and Siberia", this approach was insufficient so we ultimately decided to define a new ancestry category, "Indigenous Siberian ancestry" (HANCESTRO:0845) as the best way of representing population descriptors in this superpopulation. Work is ongoing to define matching population descriptors for some of these reference populations as very little suitable scientific documentation exists to accurately represent these populations.

# Discussion

HANCESTRO exists between the competing interests of, on the one hand, addressing the pressing concerns regarding the misrepresentation of genetic research to reinforce typological thinking and perpetuate outdated notions of race-driven biological differences, and, on the other hand, facilitating the interoperable and structured annotation of research data necessary for data discovery and reuse. The ontology presents the different descriptor types for ancestry, ethnicity and geography in parallel structures to avoid any implication of hierarchy between them. Hierarchical relationships in the ontology are used only in their strictest taxonomical sense of grouping narrower concepts under broader ones.

The ontology's framework also operationalises several of the NASEM report's recommendations. The reference population framework integrates all the reference populations provided by the International Genome Sample Resource (IGSR), including the superpopulations from the different source projects, providing stable and persistent identifiers for each population that can be used to generate machine-readable genetic similarity descriptions in line the report's best practice guidance. Following NASEM Recommendation 7 to apply labels consistently within a given context, separate groupings for geography-based descriptors and ethnicity-based descriptors allow users to apply a single descriptor type without having to implement additional filtering. HANCESTRO also uses comments to flag terms with outdated connotations or that are not in line with a community's preferred usage for self-description. In line with recommendations 2 and 6, HANCESTRO's user documentation recommends the use of multiple terms to improve clarity and acknowledge the complexity of human genetic variation, rather than enforce arbitrary single-term limitations that not only reduce the accuracy of the annotation but also erroneously reinforce the implication of clear delineations between categories.

## Uptake by the community

HANCESTRO has been adopted by a number of resources and projects beyond its original scope in the GWAS Catalog. On a technical level, the ontology, or parts of it, is imported by several downstream ontologies, including the Experimental Factor Ontology[26] (EFO), the Food Ontology[27] (FoodOn, https://foodon.org/), the Genomic Epidemiology Ontology[28] (GENEPIO, http://genepio.org/), the Proteomics Identification Database Ontology[29] (PRIDE, https://github.com/PRIDE-Archive/pride-ontology), and the African Population Ontology (AfPO). Resources that use these ontologies can, in turn, integrate the imported HANCESTRO terms into their own processes.

The ontology is also used in an increasing number of data annotation contexts, such as in the Encyclopedia of DNA Elements (ENCODE, https://www.encodeproject.org/) Portal[30], it is used in the Human Donor schema (https://www.encodeproject.org/profiles/human_donor), and the German Human Genome-Phenome Archive[31] (GHGA, https://www.ghga.de/), where it is used in a similar fashion. While the GWAS Catalog[32] does not explicitly integrate HANCESTRO identifiers in its current data model, the Catalog's curation model and HANCESTRO were developed together[6], and backwards compatibility with the original framework was maintained through HANCESTRO's development, so the two remain essentially cross-compatible. The PGS Catalog[33] reuses the GWAS Catalog ancestry framework and should therefore also be easily cross-compatible with HANCESTRO, as also indicated by its FAIRsharing record, should such an integration be required.

The CZ CELLxGENE suite of tools[34,35], whose metadata model was derived from the initial Human Cell Atlas (https://www.humancellatlas.org/) one, and the Single-Cell Pediatric Cancer Atlas[36] include HANCESTRO-based tagging for population descriptor concepts in their models. HANCESTRO annotations have also anecdotally been found in resources such as Biosamples[37] and the European Genome-Phenome Archive[38] (EGA), which leverage the common EBI ontology infrastructure, even though HANCESTRO is not an explicit part of their respective core metadata models.

The adoption of HANCESTRO across such heterogenous use cases and resources suggests that the annotation-layer design choice is generalizing beyond its original GWAS Catalog remit and meets the needs of a range of user communities.

## Design trade-offs and lessons from implementation

As previously discussed, we had to accommodate the hierarchical constraints of OWL ontologies, which make it impossible to assign a descriptory type to one term without automatically assigning the same type to its descendants. In this context, the decision was made that geography-based descriptors carry a stricter meaning than ethnicity descriptors, so it is more appropriate to allocate a geography-based descriptor into the ethnicity descriptor grouping than vice versa, as illustrated by the "Andalusian" (AfPO:0000355)/"Tunisian Andalusian" (AfPO:0000363) example.

The recognition of AfPO as an authoritative source and the associated decision to defer to AfPO for all concepts related to African populations represents another design trade-off, in particular as this choice necessitated the deprecation of existing HANCESTRO terms. Although AfPO in turn chose to import HANCESTRO's geographical framework for continents, regions and countries, their interpretation of African sub-continental regions in the context of their population definitions necessitated the re-allocation of several countries to new regions, namely South Sudan and Sudan from Northern Africa to Eastern Africa, Zambia from Eastern Africa to Central Africa and Mauritania from Western Africa to Northern Africa. AfPO's position as the authority on African populations, the extensive domain expertise of the AfPO development team and the quality of their evidence base fully justified this decision.

The more complex hierarchical structure of AfPO also affects the aforementioned nesting under the ethnicity and geography-based descriptor groupings. While these groupings are specifically intended as flat lists, with no hierarchical representations that might lead to misinterpretations, AfPO's nesting introduces some unintended nesting of ethnicity descriptors not found in the geography-based descriptors. We considered applying the ethnicity descriptor tag to leaf concepts only but this would have resulted in the exclusion of important descriptors such as "Nguni" (AfPO:0000340), a Southern Bantu population group, which includes the Ndebele (AfPO:0000350) (further divided into North Ndebele (AfPO:0000352) and South Ndebele (AfPO:00003503)), Swazi (AfPO:0000351), Xhosa (AfPO:0000116) and Zulu (AfPO:0000069) peoples. The nesting was therefore considered an acceptable trade-off to mitigate the loss of information.

## Limitations & challenges

In addition to these deliberate design choices, HANCESTRO contains a number of inherent limitations for which no solution exists or has been identified to date.

We explored different options of dealing with the representation of outdated or contentious terms in the ontology. Backwards compatibility and interoperability requirements dictated that synonyms such as “Caucasian” for “European ancestry” (HANCESTRO:0005) or “indio” for “Indigenous in South America” (HANCESTRO:0611) should be represented. An initial plan to define a dedicated annotation property for outdated synonyms was considered but was rejected as this would make it more challenging for resources implementing standard ontology usage scenarios to identify these terms correctly. Instead, these terms are flagged through the use of comments, as they are targeted at human users rather than computer systems. The downside of the synonym approach is that LLMs consuming HANCESTRO may prioritise synonyms over comments and thus give undue prominence to terms included only for historical reasons and not intended for modern use.

As mentioned in the reference population results, the SGDP-defined superpopulations for “West Eurasia” and “Central Asia and Siberia” could not be easily mapped into the existing HANCESTRO framework. While our mapping decisions were based on the evidence published by the project, it still represents an interpretation of the admixture plot that differs from the project’s own chosen cut-off. We acknowledge this limitation in our mapping and continue to monitor emerging work in this area that could inform mapping improvements.

Sparse documentation for some reference populations represents a limitation also on a wider level, making it challenging to close some of our coverage gaps for underrepresented small populations. No solution exists beyond monitoring emerging work and encouraging researchers in the fields to proactively request new terms for their populations of interest from HANCESTRO - or AfPO, in the case of African populations.

We also highlight the importance of carefully reviewing and vetting sources before incorporating new information into the ontology. The use of inappropriate references influenced by, for example, political or religious bias, could further the mischaracterisation and stigmatisation of populations that already face extensive discrimination in many areas, from research and health care to education and access to resources.

Finally, we acknowledge that HANCESTRO inherently represents a specific interpretation of population descriptors that may not align with the views and needs of every community. The Population Descriptors Working Group (PDWG) of the Polygenic Risk Methods in Diverse Populations (PRIMED) Consortium proposes an alternative approach[39] based on the view that population descriptors are inherently context-dependent. Their highly flexible data model supports the definition of population descriptors that fully meet the context needs of a project or consortium such a PRIMED. The downside of this approach is the increased variability in descriptors that may make it challenging to combine data from different sources. HANCESTRO’s global harmonised approach reduces variability but limits the accommodation of context-dependent aspects.

## Future work

Efforts are ongoing to close coverage gaps in the ontology, in particular for underrepresented populations, such as Indigenous South American, Siberian and Central Asian populations and to complete some of the existing reference populations that are not fully reflected in HANCESTRO. We continue to monitor emerging evidence that supports the definition of new descriptors and welcome requests from the research community for new terms. We also continue our collaboration with AfPO, referring new term and change requests for African populations to them, and ensuring new releases in either ontology are picked up by the other in a timely fashion.

While we endeavour to find supporting documentation in the scientific literature for defining population descriptors, we did not include definition sources directly in the ontology until recently. These gaps will be closed as much as possible over time, using the evidence standards set by AfPO.

We also continue to improve HANCESTRO's alignment with the recommendations of the NASEM report, including exploring alternative terminology such as renaming "ancestry category" to "ancestry label" to further reduce any chance of misunderstanding. We are also exploring alternative strategies for capturing outdated language beyond the current synonym/comment compromise.

A recurring issue in the development of HANCESTRO is its small team size and limited resources. Exploring scientific literature to identify appropriate sources is a laborious task, while even the ODK release framework still requires manual checking steps. For this reason, we are investigating the use of agentic AI workflows such as those already in place in the Experimental Factor Ontology[26] (EFO), Monarch Disease Ontology[40] (MONDO) and Gene Ontology[41] (GO). These workflows could replace many of the more time consuming aspects of maintaining HANCESTRO, including closing the gaps in supporting documentation, thus allowing the curators to focus on validation and quality assurance.

As HANCESTRO's uptake by the community grows, we are also considering putting in place more formal mechanisms to periodically reevaluate the use of concepts in HANCESTRO in line with evolving scientific and social understanding of population descriptors. A rigorous governance framework supported by external advisors would greatly strengthen HANCESTRO's utility and relevance as a community resource.

## Conclusion

HANCESTRO occupies a deliberately narrow role: it does not classify human populations, nor does it adjudicate which population descriptor is appropriate for a given study. Instead, it provides a shared, machine-readable vocabulary for annotating the descriptors researchers and data generators have already chosen to report, whether self-identified ethnicity, geographic origin, or genetic similarity to a named reference population. This annotation-first design does not resolve the deeper tensions identified by the NASEM report, nor does it aim to; rather, it seeks to ensure that whatever descriptors the field ultimately converges on can be recorded, queried, and reused consistently. As best practices in this area continue to evolve, HANCESTRO's own structure and scope will need to evolve alongside them — a responsibility we intend to meet through the same community-driven, transparent development process described in this paper.

# Resource availability

## Lead contact

Requests for further information and resources should be directed to and will be fulfilled by the lead contact, Danielle Welter (danielle.welter@lnds.lu).

## Materials availability

This work did not generate new materials.

## Data and code availability

The HANCESTRO source files, scripts and documentation are licensed under CC-BY 4.0 and openly available from the GitHub repository https://github.com/EBISPOT/hancestro.

# Acknowledgements

The authors would like to acknowledge the HANCESTRO user community, in particular the contributors who supported the evolution of the ontology by submitting requests for new terms, identifying issues and suggesting improvements. Special thanks go to Prof Alia Benkahla, AfPO lead, Institut Pasteur de Tunis, Tunisia, for her invaluable advice and expertise on African populations; Dr Laura Harris, GWAS Catalog Coordinator at EMBL-EBI, for her advice and input on improving alignment with NASEM recommendations; Prof Nicola Mulder, University of Cape Town, for her advice and guidance; Dr Norbert Tavares, Senior Program Officer at the Chan Zuckerberg Initiative, for his advice and support in shaping the project and reaching out to new contributors.
This project has been made possible in part by grant number 2023-325293 from the Chan Zuckerberg Initiative DAF, an advised fund of Silicon Valley Community Foundation. HP, ZMP and AC were funded in part by EMBL-EBI Core Funds.

# Author contributions

Authors' contributions are listed as per Contributor Roles Taxonomy (https://credit.niso.org). DW, Conceptualization, Data Curation, Methodology, Project administration, Resources, Writing – original draft, Writing – review & editing; AC, Methodology, Resources, Software, Writing – review & editing; ZMP Data Curation, Methodology, Resource, Writing - review & editing; MC, Data Curation, Resources, Writing - review & editing; HP, Conceptualization, Project administration, Resources, Writing – review & editing.

# Declaration of interests

The authors declare that they have no competing interests.

# Declaration of generative AI and AI-assisted technologies in the writing process

During the preparation of this work, the authors used Claude in order to enhance the structure and clarity of some of the sections. After using this tool, the authors reviewed and edited the content as needed, and take full responsibility for the content of the published article.